\documentstyle[12pt,chicagob]{article}
\input{psfig}
\newtheorem{THEOREM}{Theorem}[section]
\newenvironment{theorem}{\begin{THEOREM} \hspace{-.85em} {\bf :} }%
                        {\end{THEOREM}}
\newtheorem{LEMMA}[THEOREM]{Lemma}
\newenvironment{lemma}{\begin{LEMMA} \hspace{-.85em} {\bf :} }%
                      {\end{LEMMA}}
\newtheorem{COROLLARY}[THEOREM]{Corollary}
\newenvironment{corollary}{\begin{COROLLARY} \hspace{-.85em} {\bf :} }%
                          {\end{COROLLARY}}
\newtheorem{PROPOSITION}[THEOREM]{Proposition}
\newenvironment{proposition}{\begin{PROPOSITION} \hspace{-.85em} {\bf :} }%
                            {\end{PROPOSITION}}
\newtheorem{DEFINITION}[THEOREM]{Definition}
\newenvironment{definition}{\begin{DEFINITION} \hspace{-.85em} {\bf :} \rm}%
                            {\end{DEFINITION}}
\newtheorem{CLAIM}[THEOREM]{Claim}
\newenvironment{claim}{\begin{CLAIM} \hspace{-.85em} {\bf :} \rm}%
                            {\end{CLAIM}}
\newtheorem{EXAMPLE}[THEOREM]{Example}
\newenvironment{example}{\begin{EXAMPLE} \hspace{-.85em} {\bf :} \rm}%
                            {\end{EXAMPLE}}
\newtheorem{REMARK}[THEOREM]{Remark}
\newenvironment{remark}{\begin{REMARK} \hspace{-.85em} {\bf :} \rm}%
                            {\end{REMARK}}

\newcommand{\thm}{\begin{theorem}}
\newcommand{\lem}{\begin{lemma}}
\newcommand{\pro}{\begin{proposition}}
\newcommand{\dfn}{\begin{definition}}
\newcommand{\rem}{\begin{remark}}
\newcommand{\xam}{\begin{example}}
\newcommand{\cor}{\begin{corollary}}

\newcommand{\ethm}{\end{theorem}}
\newcommand{\elem}{\end{lemma}}
\newcommand{\epro}{\end{proposition}}
\newcommand{\edfn}{\bbox\end{definition}}
\newcommand{\erem}{\bbox\end{remark}}
\newcommand{\exam}{\bbox\end{example}}
\newcommand{\ecor}{\end{corollary}}

\newcommand{\beqn}{\begin{equation}}
\newcommand{\eeqn}{\end{equation}}

\newcommand{\bbox}{\vrule height7pt width4pt depth1pt}

\newcommand{\clm}{\begin{claim}}
\newcommand{\eclm}{\end{claim}}

\renewcommand{\phi}{\varphi}

\newcommand{\G}{{\cal G}}

\newcommand{\R}{{\cal R}}

\newcommand{\ie}{i.e.,~}

\newcommand{\ol}{\setlength{\itemsep}{0pt}\begin{enumerate}}
\newcommand{\eol}{\end{enumerate}\setlength{\itemsep}{-\parsep}}
\newcommand{\ul}{\setlength{\itemsep}{0pt}\begin{itemize}}
\newcommand{\dl}{\setlength{\itemsep}{0pt}\begin{description}}
\newcommand{\edl}{\end{description}\setlength{\itemsep}{-\parsep}}
\newcommand{\eul}{\end{itemize}\setlength{\itemsep}{-\parsep}}

\newcommand{\commentout}[1]{}

\newcommand{\bi}{\begin{itemize}}
\newcommand{\ei}{\end{itemize}}
\newcommand{\be}{\begin{enumerate}}
\newcommand{\ee}{\end{enumerate}}

\newcommand{\Gz}{\G_0}

\include{epsf}
\renewcommand{\omega}{w}

\begin{document}

\newcommand{\strat}{{\bf s}}
\pagenumbering{arabic}
\title{A Computer Scientist Looks at Game Theory}
\author{Joseph Y.\ Halpern%
\thanks{This paper is based on an invited talk I gave at Games 2000 in
Bilbao, Spain.  Supported in part by NSF under grants IRI-96-25901 and
IIS-0090145 and by ONR under grants  N00014-00-1-03-41,
N00014-01-10-511, and N00014-01-1-0795.}\\ 
   Cornell University\\
   Computer Science Department\\
   Ithaca, NY 14853\\
   halpern@cs.cornell.edu\\
   http://www.cs.cornell.edu/home/halpern
}
\maketitle

\begin{abstract} I consider issues in distributed computation that
should be of relevance to game theory.  In particular, I focus on (a)
representing knowledge and uncertainty, (b) dealing with failures, and (c)
specification of mechanisms.
{\em Journal of Economic Literature\/} Classification Numbers: D80,
D83.
\end{abstract}

\section{Introduction}
There are many areas of overlap between computer science and game
theory.  The influence of computer science has been felt perhaps most
strongly through complexity theory.   Complexity theory has been viewed
as a tool to help capture bounded rationality, going back to the work of
Neyman \citeyear{Ney85} and Rubinstein \citeyear{Rub85}.  In addition,
it is well understood that complexity-theoretic notions like
NP-completeness help categorize the intrinsic difficulty of a problem.
Thus, for example, a result showing that, even in simple settings,
the problem of optimizing social welfare is NP-hard \cite{KMT00} shows
that the standard procedure of applying Clarke's mechanism, despite its
many benefits, is not going to work in large systems.  

Perhaps less obvious is the interplay between game theory and work in 
distributed computing.  At the surface, both areas are interested in
much the same problems: dealing with systems where there are many
agents, facing uncertainty, and having possibly different goals.  
In practice, however, there has been significant difference in 
emphasis in the two areas.  In distributed computing, the focus has been
on problems such as fault tolerance, scalability, and proving
correctness of algorithms; in game theory, the focus has been on
strategic concerns (that is, playing so as to optimize returns, in light
of the preferences of other agents).  
In this paper, I hope to make the case that each
area has much to learn from the other.
I focus on three particular topics: 
\begin{itemize}
\item the representation of games (and, in
particular, the knowledge and uncertainty of players in a game),  
\item strategic concerns vs.~fault tolerance, and 
\item specification of mechanisms.
\end{itemize}
The following sections deal with each of these topics in turn.

\section{Representing Games as Systems}
\begin{figure}[htb]
\begin{center}
\begin{tabular}[b]{ll}
\begin{tabular}[b]{|| l | l  l || c}
\hline
{ } & A  & D \\ 
\hline 
aa  &(3,3)  & (4,2)\\
ad  & (3,2) & (4,2) \\
da  & (1,2) & (1,3) \\
dd  &  (1,2) & (1,3) \\
\hline
\end{tabular}
&\epsfysize=7.0cm \epsffile{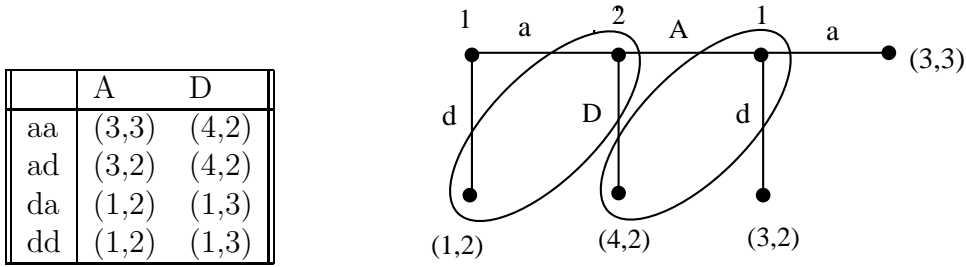}
\end{tabular}
\end{center}
\caption{Representing a game in normal form and in extensive form.}
\label{gamerep}
\end{figure}

In order to analyze a game, we must first represent it.  The two
most common representations in the literature are the normal-form
representation and the extensive-form representation.  
As is well known, the extensive-form representation brings out the
temporal aspects of the game better, as well as explicitly representing
(at least some aspects) of the players' knowledge.
Consider the game that is  represented in Figure~\ref{gamerep} in both
normal form and extensive form.
The extensive-form representation  brings out clearly that the 
game takes place over time, with the first player's second move, for
example, occurring after the second player's first move.  Moreover, when
the first player makes that second move, he does not know what the
second player's move is.  However, as is also well known, the
information sets used in the extensive-form representation do not
capture all aspects of a player's information.  For example, they cannot
be used to capture beliefs one player has about what strategy the other
player is using, or notions like rationality and common knowledge of
rationality.   The state-space representation does better in this regard.

\subsection{The state-space representation}

The state-space representation, first used in the economics literature
by Aumann \citeyear{Au},  is actually a variant of the standard
possible-worlds model for 
knowledge in the philosophical literature that goes back to Hintikka
\citeyear{Hi1}; see \cite[Section 2.5]{FHMV} for discussion.  
In this representation, each state in a state space $\Omega$ is a complete
description of the world, which includes what happened in the past and
what will happen in the future, the agents' beliefs and knowledge, and so
on.   

\begin{figure}[htb]
\centerline{\psfig{figure=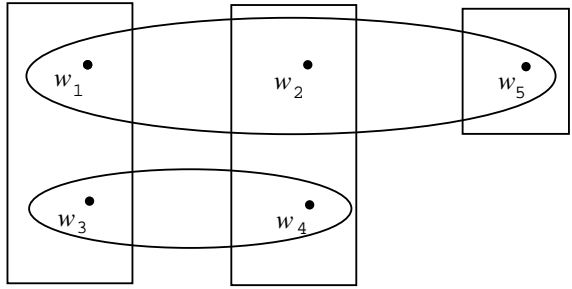,height=1.5in}}
\caption{Representing the game of Figure~\ref{gamerep} using a state space.}
\label{gamerep1}
\end{figure}

One representation of the game in
Figure~\ref{gamerep} using a state space is given in
Figure~\ref{gamerep1}.  
Let $\Omega = \{\omega_1, \ldots, \omega_5\}$.
With each state $\omega \in \Omega$ is associated the strategy profile
$\strat(\omega)$ played at $\omega$.  
In this example, 
\begin{itemize}
\item $\strat(\omega_1) = \strat(\omega_5) = (aa,A)$
\item $\strat(\omega_2) = (aa,D)$
\item $\strat(\omega_3) = (ad,A)$
\item $\strat(\omega_4) = (ad,D)$
\end{itemize}
In addition, there are two partitions associated with this state space,
one for player 1 (denoted by ellipses in Figure~\ref{gamerep1}) and one
for player 2 (denoted by rectangles).  The fact that $\omega_3$ and
$\omega_4$ are both in the same cell of player 1's partition means that
player 1 can't tell, in state $\omega_3$, if the actual state is
$\omega_3$ or $\omega_4$.  Note that in every cell for player 1, player
1 is following the same strategy; similarly for player 2.  This is meant
to capture the intuition that the players know their strategy.  Further
note that not all strategy profiles are associated with a state (for
example, $(dd,A)$ is not associated with any state) and some profiles
(such as $(aa,A)$ in this case) can be associated with more than one
state.  There is more to a state than the strategy profile used there.
For example, in $\omega_5$, player 2 knows that the strategy profile is
$(aa,A)$, while in $\omega_1$, player 2 considers it possible that $(ad,A)$ is played.

The state-space representation suffers from some of the same problems as
the normal-form representation.
While it does a reasonably good job of
capturing an agent's knowledge, it does not do such a good job of
describing the play of the game---who moves when, and what the possible
moves are.  Moreover, because time is not explicit in this representation, it
becomes difficult to model statements such as ``I know now that after I
move my opponent will not know \ldots''.  More seriously, I would claim,
neither the state-space representation nor the extensive-form
representation makes it  
clear where the knowledge is coming from.  Exactly what does it mean
put two nodes or two states in the same information set?

\begin{figure}[htb]
\centerline{\psfig{figure=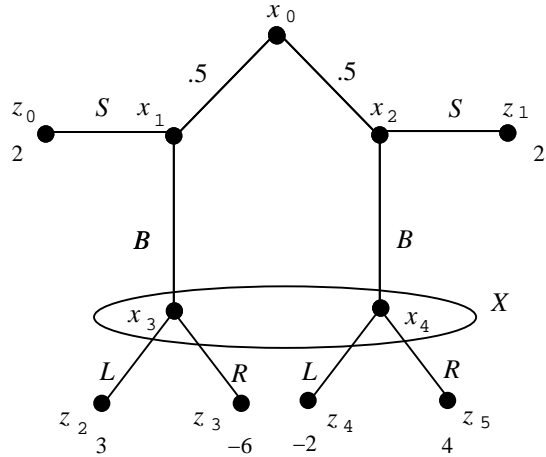,height=2.5in}}
\caption{A game with imperfect recall.}
\label{fig2}
\end{figure}

This issue becomes particularly relevant when considering games with
imperfect recall.  Considering the single-agent game described in
Figure~\ref{fig2}, introduced by Piccione and Rubinstein
\citeyear{PR97}. It is a game of imperfect recall since at the information set
$\{x_3,x_4\}$, the agent has forgotten nature's initial move (i.e.,
whether it was earlier at $x_1$ or $x_2$).

It is not hard to show that the strategy that maximizes expected utility
in this game chooses action $S$ at node $x_1$, action $B$ at node
$x_2$, and action $R$ at the information set $X$ consisting of $x_3$ and
$x_4$.  Call this strategy $f$.  Let $f'$ be the strategy of choosing
action $B$ at $x_1$, action $S$ at $x_2$, and $L$ at $X$.
Piccione and Rubinstein argue that if node $x_1$ is reached, the
agent should reconsider, and decide to switch from $f$ to $f'$.
{\em If the agent is able to remember that he switched strategies},
then this is correct; the
agent is indeed better off (under any reasonable notion of ``better
off'') if he switches.

The reason for the time inconsistency here is that an agent's
strategy must dictate the same action at nodes $x_3$ and $x_4$, since
they are in the same information set.  Intuitively, since the agent cannot
distinguish the nodes, he must do the same thing at both.  If
the agent had perfect recall, he could distinguish the nodes.
The optimal strategy with perfect recall amounts to 
switching from $f$ to $f'$ at $x_1$:
the agent plays $L$ at $x_3$ (as he would with $f'$) and $R$ at $x_4$
(as he would with $f$).  However, by having the ability to 
remember that he has switched strategies, the agent is able to simulate
perfect recall.  If he is using $f'$ at the information set, he knows he
must have been at $x_1$ (and thus is at $x_3$); similarly, if he is
using $f$ at the information set, then he must be at $x_4$.
What does it mean, then, to put $x_3$ and $x_4$ in the
same information set?  What entitles a modeler to put an ellipse around
$x_3$ and $x_4$?

In the computer science literature  a different approach is used to
represent knowledge in multi-agent systems.  This approach goes back to
\cite{HF87,HM1}, and has been used quite  successfully to model
distributed systems applications \cite[Chapters 4--7]{FHMV}.  
The approach can be viewed as combining features of both game trees and
the state-space representation.   Not surprisingly, it can also be used
to model games.    The idea is that a game is
represented as a multi-agent {\em system}.  In the description of the system, 
the actual play of the game is distinguished from what goes on in the
agent's mind.  I claim that doing so can clear up the type of problems
encountered in the game in Figure~\ref{fig2}.  In the remainder of this
section, I describe the approach and sketch how it can be used to deal
with this game.  I describe some further advantages of the approach in
the next section.

\subsection{The multi-agent systems approach}

The basic framework is easy to describe although, unfortunately, the
description requires a number of definitions.  I review the relevant
definitions in this section.%
\footnote{The following description is taken almost verbatim from \cite{Hal15}.
See \cite[Chapter 4]{FHMV} for more details.}
To describe the agent's state of mind,
we assume that, at every point in time, the agent is in some {\em
state}.  Occasionally this is called a {\em local state}, to
distinguish it from a {\em global state}, which is defined below.
The local state is essentially what is called an agent's {\em type\/}
in the game theory literature.
Intuitively, it encapsulates all the information to
which the agent has access.  Deciding how to model the state can be
quite a nontrivial issue.  In a poker game, a player's state might
consist of the cards he currently holds, the bets made by the other
players, other cards he has seen, and whatever information he has about
the strategies of the other players.  A forgetful player
may not remember all the details of the bets made by the other players;
his state would reflect this.

To describe the external world, we use an {\em
environment}, which is also in some state at every point in time.
Roughly speaking, the environment's state describes everything relevant
to the system that is not part of the agents' states.
For example, when describing a game, we can take
the environment's state at a given point to consist of
the sequence of actions that have been performed up to that point; in
addition, at points representing the end of play, the environment's
state can include the payoffs.
If we do this, we can essentially identify the possible environment
states with the nodes in the game tree.

The configuration of the system as a whole can be described by a {\em
global state}, a tuple of the form $(\ell_e, \ell_1, \ldots, \ell_n)$,
where $\ell_e$ 
is the environment's state, and $\ell_i$ is agent $i$'s state, $i = 1,
\ldots, n$.  A global state describes the system at a given point in
time.  We are typically interested in dynamic systems that change over
time.  A {\em run\/} is a function from time (which is taken for
simplicity to range over the natural numbers) to global states.
Intuitively, a run is a complete description of how the system's global
state evolves over time.  For example, when analyzing a game, a run
could be a particular play of the game.
Thus, if $r$ is a run, $r(0)$ describes the initial global state of the
system, $r(1)$ describes the next global state, and so on.
A {\em point\/} is a pair $(r,m)$ consisting of a run $r$ and time $m$.
If $r(m) =
(\ell_e,\ell_1, \ldots, \ell_n)$, let $r_i(m) = \ell_i$.  Thus, $r_i(m)$
is agent $i$'s local state at the point $(r,m)$.

Finally, a {\em system\/} is formally defined to be a set of runs.
Intuitively, a system is being identified with its set of possible
behaviors.  Thus, for example, the game of bridge can be identified with
all the possible games of bridge that can be played
(where a run describes a particular play of the 
game, by describing the deal of the cards, the bidding, and the play of
the hand).

Notice that information sets are conspicuously absent from this
definition.  Information sets in fact do
not have to be specified exogenously; they can be reconstructed from the
local states.  Given a system, that is, a set $\R$ of runs,
we can define an equivalence relation on the points in $\R$.
The point $(r,m)$ is {\em indistinguishable from $(r',m')$ by agent $i$},
denoted $(r,m) \sim_i
(r',m')$, if $r_i(m) = r'_i(m)$.  Thus, two points are indistinguishable
by agent $i$ if agent $i$ has the same local state at both
points.  Clearly $\sim_i$ is an equivalence relation.
The $\sim_i$ relations can be viewed as defining information sets.
However, note that even a point where agent $i$ does not move is in
an information set for agent $i$.  We can now define what it means for
agent $i$ to {\em know\/} an event $E$: agent $i$ knows $E$ at a point $(r,m)$
if the set of points indistinguishable from $(r,m)$ by agent $i$ (that
is, $\{(r',m'): (r,m) \sim_i (r',m')\}$) is a subset of $E$.

A {\em protocol\/} for an agent in this setting is a function
from that agent's local states to actions (or to a distribution over
actions, if we want to consider randomized protocols).  
Essentially, a protocol is just a strategy; a randomized protocol is
essentially a behavior strategy. The definition captures the 
intuition that what an agent does can depend only on her information
(i.e., her local state). 
At two points where the agent has the same information, the
agent must do the same thing.  The agent's local state in this setting
corresponds to the agent's information set in the extensive-form
representation of the game.  However, thinking in terms of local states 
seems more natural for protocol designers than thinking in terms of
information states in a game tree.   It is much more natural to write a
program that says ``if you have received a message then send an
acknowledgment'' than to describe the whole interaction (i.e., the game
tree), put an ellipse around those nodes in the game tree in which the
agent has received the message (this would be the information set
corresponding to the local state where the agent has received the
message), and describe the strategy that performs the action of
acknowledging the message at that information set.  Most importantly, by
using local states, it becomes clear exactly what an agent's information
is at any point and, thus, what the agent's information set should be.

We often think of systems as generated by agents running a {\em joint
protocol}, that is, a tuple consisting of a protocol for each agent.
Intuitively, starting in some initial global state(s), we run the joint
protocol and see what happens step by step.  But what exactly  happens
when a joint protocol $P$ is run?  That depends on 
the setting, or {\em context\/}, in which $P$ is being run.  The context
determines, among other things, what the environment does.  The environment
is viewed as running a protocol just like the agents; 
its protocol is used to capture features of the setting 
such as ``all messages are delivered within 5 rounds'' or 
``messages may be lost''.  Roughly speaking, the environment's protocol
corresponds to ``nature's strategy''---the way nature plays the game.  
The context also determines how the actions performed by the protocol
change the global state.  
Formally, a context $\gamma$ is a tuple $(P_e,\Gz,\tau)$,
where $P_e$ is a protocol for the environment,
$\Gz$ is a set of initial global states (intuitively, the set of states
in which it is possible to start a run), and
$\tau$ is a {\em transition function}.%
\footnote{Often it is also convenient to include in the tuple a
component describing which runs are admissible.  This is done in
\cite{FHMV} to capture notions such as {\em fairness}: if a message is sent
infinitely often, it is eventually received.  (Thus, runs where a
message is sent infinitely often but not received are considered
inadmissible.)  Since admissible runs play no role in the discussion
here, I omit this component from the context.}
The transition function $\tau$ captures the
effect of actions; formally, it describes how the actions performed by
the agents and the environment change the global state by associating
with each {\em joint action\/} (a tuple consisting of an action for the
environment and one for each of the agents)
a {\em global state transformer}, that is, a mapping from global states 
to global states.  

For ease of exposition, suppose that $P$ is a deterministic joint protocol.
A run $r$ is {\em consistent with $P$ in context
$\gamma = (P_e,\Gz,\tau)$\/} if 
the initial global state $r(0)$ is one of
the global states in $\Gz$,
and for all $m$, the transition
from global state $r(m)$ to $r(m+1)$ is the result of performing the
joint action specified by~$P$ and the environment protocol $P_e$
in the global state $r(m)$.
A system $\R$ {\em represents\/} a joint protocol $P$ in a context
$\gamma$ if it consists of all runs in $\Psi$ consistent with $P$ in
$\gamma$.  (If $P$ is a randomized protocol, essentially the same
construction gives the set of runs consistent with $P$ together with
a probability distribution on them; I omit the formal details here.)

The role of the context should become clearer in the examples in
Section~\ref{sec:failures}.

\subsection{From game trees to systems}
There is a great deal of flexibility in representing a game using a
system.  It depends on what the local states are.  One possibility
essentially directly emulates the extensive-form representation.  In
this approach, each run in the system correspond to a play of the
game---i.e., a branch in the game tree.  Thus, there is essentially one
run for each terminal node in the game tree.  Under this representation,
the environment state at a certain point is the node in the game tree
(or, equivalently, the sequence of actions taken to reach that node);
the environment state at points that correspond to terminal nodes would
also include the payoff.  An agent's local state could then simply be
his information set.
A more natural representation of an agent's local state might be the
sequence of actions she recalls seeing.%
\footnote{Some decision also has to be made as to the agent's local
state at nodes where the agent does not move.  There are a number of
reasonable choices; the one made does not affect the main points.}

Note that if we represent a game this way, there is no information about
strategies, just as there is no information about strategies in the
extensive-form representation.  This is not due to a lack of expressive
power in the systems framework; rather, it is due to the choice of local states.

Another choice is closer to the state-space representation.  Each state
in a state space corresponds to a run in the system.  The play of the
game in the run is the play generated by the strategy profile associated
with the state. Again, the environment state could be the node reached
(and the payoff, if it is a terminal node).  But now an agent's local
states would include a representation of the strategy she is using, what
she recalls having seen thus far, and some representation of her beliefs
about other agents' strategies.    If it is common knowledge that agents
do not switch strategies in the course of the game, this common
knowledge can be  represented by considering systems that consist only
of runs where the players strategy does not change over time.  

What happens if agents can switch strategies?  Again, there is no
difficulty modeling this in the framework.  (But note that, strictly
speaking, switching strategies should then be considered one of the
actions in the game.)  An agent's local state would then include her
current strategy (or perhaps the sequence of strategies she has used up
the current time.)  If we model the game in Figure~\ref{fig2}
using such a  system, if the player knows that he will switch from $f$ to
$f'$ at $x_1$, then at points in the system corresponding $x_3$, he 
will know that he is at $x_3$ (because, according to his local state, he
is using strategy $f'$), while at $x_4$, he will know that he is at
$x_4$.  If falls right out of the representation that agents that
are allowed to switch strategies and know their current strategy will be
able to simulate perfect recall.

The key point is that the use of local states in the runs and systems
framework  forces a modeler to be explicit about what an agent knows and does
not know in  a way that drawing ellipses in the extensive-form
representation or the state-space representation does not.
This, in turn, can force some important discipline on the modeler of the game.
In the game in Figure~\ref{fig2}, for example, the modeler is forced to 
to say whether the player allowed to switch strategies and, if so,
whether he keeps track of his current strategy.  The answer to this
question is modeled in the player's state.  Whatever the answer to the
question, there will be no time inconsistency. 
(See \cite{Hal15} for a more detailed discussion of this example and the
notion of modeling games using the systems framework.)  

\section{Coping with Failures and Asynchrony}\label{sec:failures}

There is a great deal of work in the distributed systems  literature on
designing protocols to deal with certain paradigmatic problems.  There
is a lot of overlap in spirit between this work and much of the work in
the game theory literature on mechanism design.  There are, however,
also significant differences.  
Game theory focuses on autonomous agents and their strategic interests.
In the distributed systems literature, the ``agents'' are processes,
which are given a protocol to run by a systems designer.   
The distributed systems literature focuses on what can go
wrong and what makes running the protocols difficult---communication
failures, process failures, asynchrony, and the complexity issues involved
in dealing with large systems.  All of these  issues are, 
by and large, not discussed in the game theory literature.  
In this section, I give examples of problems in which issues of failures
and asynchrony arise.   These examples also illustrate some
other advantages of using the systems representation.

\subsection{Coordinated Attack}\label{s:coord}
The coordinated
attack problem is a well-known problem from the distributed systems
folklore \cite{Gray}.  The following description of the problem is taken
from \cite{HM1}; the discussion of it is taken from \cite{Hal32}.
\begin{quote}
Two divisions of an army are camped on two hilltops overlooking
a common valley.  In the valley awaits the enemy.  It is clear that
if both divisions attack the enemy simultaneously they will win
the battle, whereas if only one division attacks it will be defeated.
The generals do not initially have plans for launching an attack
on the enemy, and the commanding general of the first division wishes to
coordinate a simultaneous attack (at some time the next day).  Neither
general will decide to attack unless he is sure that the other will
attack with him.  The generals can only communicate by means of a
messenger.  Normally, it takes the messenger one hour to get from one
encampment to the other.  However, it is possible that he will get lost
in the dark or, worse yet, be captured by the enemy.  Fortunately,
on this particular night, everything goes smoothly.  How long will it
take them to coordinate an attack?
\end{quote}

In the language of game theory, the problem here is to design a
mechanism that guarantees that the generals coordinate, despite the
possibility of messages being lost.  As is typically the case in
distributed systems problems, there is no discussion of what the 
payoff is for general $A$ and $B$ if both attack, neither does, or one
does and the other does not.  Nor is there is any mention of
probabilities (in particular, the probability that the messenger will
arrive).  While interesting issues certainly arise if strategic concerns and
probability are added 
(see, for example, Rubinstein's \citeyear{Rub89} lovely results),
there are good reasons why these issues are being ignored here.
The coordinated attack problem is an attempt to understand the effect 
of possible communication failures on coordination.
The generals are viewed as being on the same ``team'', with identical
utilities, playing against ``nature'' or the ``environment'', which
controls communication. 
We could capture the intuition behind the problem by giving each general
payoff $L$ if they do not coordinate, utility $M$ if neither attacks,
and utility $H$ if both attack, with $L < M < H$, but no new issues
would arise if we did so.  The real interest here is not in the
strategic behavior of the generals, but whether they can achieve
coordination when playing against nature.  

We could also add a
probability that a message arrives.  The problem is that, for the
situations which the coordinated attack problem was intended to
abstract, it is often quite difficult to characterize this probability.
For example, one reason that messages fail to arrive in real systems is
message congestion, often caused by ``hotspots''.  The probability of
message congestion is extremely difficult to characterize.  

Turning to the analysis of the problem,
suppose that the messenger sent by General $ A $ makes it to General $ B $
with a message saying ``Let's attack at dawn.''  Will General
$ B $ attack?  Of course not, since  $ A $
does not know that $B$ got the message, and thus may not attack.
So  $ B $ sends the messenger back with an acknowledgment.
Suppose the messenger makes it.  Will  $ A $ attack?
No, because now  $ B $ does not know that  $A$
got the message, so  $B$
thinks  $ A $ may think that he ($ B $) didn't get the original
message, and thus not attack.  So $ A $ sends the messenger back
with an acknowledgment.  But of course, this is not enough either.
 
In terms of knowledge, each time the messenger makes a transit, the {\em
depth\/} of the generals' knowledge increases by one.  More precisely,
let $E$ be the event ``a message saying `Attack at
dawn' was sent by General $A$''.  When General $B$ gets the message, $B$
knows $E$. When $A$ gets $B$'s acknowledgment, $A$ knows that $B$ knows
$E$.  Every pair of subsequent acknowledgment leads to one more level
of ``$A$ knows that $B$ knows.''  However, although more
acknowledgments keep increasing the depth of knowledge, it is not hard
to show that by following this protocol, the generals never attain {\em common
knowledge\/} that the attack is to be held at dawn, where common knowledge
describes the event that $A$ knows that $B$ knows that $A$ knows that
$B$ knows {\em ad infinitum}.
 
What happens if the generals use a different protocol?
That does not help
either.  As long as there is a possibility that the messenger may get
captured or lost, then common knowledge is not attained, even if the
messenger in fact does deliver his messages.  It would take us too far
afield here to completely formalize these results (see \cite[Section
6.1]{FHMV} for 
details), but it is not hard to give a rough description.  A {\em
context $\gamma$ displays unbounded message delays (umd)\/} if, roughly
speaking, for all systems $\R$ that represent a protocol $P$ run in
context $\gamma$, runs 
$r \in \R$, and agents $i$, if $i$ receives a
message at time $m$ in $r$, then for all $m' > m$, there is another run
$r' \in \R$ that is identical to $r$
up to time $m$ except that agent $i$ receives no messages in $r'$
between times $m$ and $m'$ inclusive, and no agent other than possible
$i$ can distinguish $r$ and $r'$ up to time $m'$ (\ie $r_j(m'') =
r'_j(m'')$ for $m'' \le m'$ and $j \ne i$).  	
That is, $r'$ looks the same as $r$ up to time $m'$ to each agent except
possibly $i$, and all messages that $i$
receives in $r$ between times $m$ and $m'$ are delayed until after time
$m'$ in $r'$.  
We can think of umd as characterizing a property of the environment's 
protocol in context $\gamma$.  Intuitively, it is the environment that
decides whether or not a message is delivered; in a context with umd,
the environment is able to hold up messages for an arbitrary amount of
time.  
 
\thm\label{coord1} {\rm \cite{HM1}} If context $\gamma$ displays
umd and $\R$ is a 
system that represents some protocol $P$ in context $\gamma$, then at no
point in $\R$ can it be common knowledge that a message has been
delivered. \ethm
 
This says that, in a context that displays umd, 
no matter how many messages arrive, the generals cannot attain
common knowledge that any message whatsoever has been delivered.  Since
it can never become 
common knowledge that a message has been delivered, and message delivery
is a prerequisite for attack, it is not hard to show that it can never
become common knowledge among the generals that they are attacking.
More precisely, let {\em attack\/} be the event that consists of the
points where both generals attack.
 
\cor\label{needck}  If context $\gamma$ displays umd and $\R$ is a 
system that represents some protocol $P$ in context $\gamma$, then at no
point in $\R$ can {\em attack\/} be common knowledge among the generals.
\ecor

Why is it relevant that the generals can never get common knowledge of
the fact that they are attacking?  Our interest here is not common
knowledge, but coordinated attack.  What does common knowledge have to
do with coordinated attack?  As the next result shows, a great deal.
Common knowledge is a prerequisite for coordination.  
Let a {\em system for coordinated attack\/} be one that represents a
protocol for coordinated attack. 
 
\thm\label{nock} {\rm \cite{HM1}} In a system for coordinated attack,
when the generals attack, {\em attack\/} must be common knowledge among
the generals.
\ethm

The statement of the coordinated attack problem assumes that 
the generals have no initial plans for attack.  This can be formalized by
assuming  that, in the absence of messages, they will not attack.  
With this assumption, Corollary~\ref{needck} and Theorem~\ref{nock}
together give the following result. 
\cor\label{nocoord} If context $\gamma$ displays umd and $\R$ is a 
system that represents some protocol for coordinated attack in context
$\gamma$, then at no point in $\R$ do the generals attack. \ecor
Note that this result can be expressed in game theoretic terms: it is
impossible to design a mechanism that guarantees coordinated attack.
These results show not only that coordinated attack is impossible (a fact
that was well known \cite{YC}), but {\em why\/} it is impossible.
The problem is due to a combination of (1) the unattainability of common
knowledge in certain contexts and (2) the  need for common knowledge to
perform coordination.  

It is worth stressing the role of systems and contexts in stating these
results.  The notion of ``communication not being guaranteed'' was
formulated in terms of a condition on contexts (umd).
Theorem~\ref{coord1} show that common knowledge in any system that can
be generated in a context satisfying umd.  The need for common knowledge
to coordinate is also formulated in terms of systems.  
The framework of runs and systems is well suited to formulating these
conditions.

\subsection{Byzantine Agreement}
The coordinated attack problem focused on communication problems.  
{\em Byzantine agreement\/} is another paradigmatic problem in the
distributed systems literature; it brings out issues of process
failures as well as asynchrony.    In this problem, there are assumed to
be $n$ soldiers, up to $t$ of which may be faulty (the $t$ stands for {\em
traitor}); $n$ and $t$ are assumed to be common knowledge.
Each soldier starts with an initial preference, to either attack or
retreat. (More precisely, there are two types of nonfaulty
agents---those that prefer to attack, and those that prefer to retreat.)
We want a protocol (i.e., a mechanism) with the following properties:
\begin{itemize}
\item All {\em nonfaulty\/} soldiers reach the same decision.
\item If all the soldiers are nonfaulty and their initial preferences
are identical, then the final decision agrees with their initial
preferences.%
\footnote{This condition simply prevents the obvious trivial solutions,
where the soldiers attack no matter what, or retreat no matter what.
Similarly, the statement ``The generals do not initially have plans to
attack'' in the description of the coordinated attack problem is
implicitly meant to prevent a similar trivial solution in the case of
coordinated attack.}
\end{itemize}

This problem has been studied in detail.  There have been literally
hundreds of papers on Byzantine agreement and closely related topics.
The problem 
was introduced by Pease, Shostak, 
and Lamport \citeyear{PSL}; Fischer \citeyear{Fisbyz} gives an overview
of the state of the art in the early 1980's; Linial \citeyear{Linial94}
gives a more recent discussion; Chor and Dwork \citeyear{CD89} survey
randomized algorithms for Byzantine agreement.  Whether the Byzantine
agreement 
problem is solvable depends in part on what types of failures are
considered, on whether the system is {\em synchronous\/} or {\em
asynchronous}, and on the ratio of $n$ to $t$.  Roughly speaking, a system
is synchronous if there is a global clock and agents move in lockstep; a
``step'' in the system 
corresponds to a tick of the clock. In an asynchronous system, there is
no global clock.  The agents in the system can run at arbitrary rates
relative to each other.  One step for agent 1 can correspond to an
arbitrary number of steps for agent 2 and vice versa.  Synchrony is an
implicit assumption in essentially all games.  Although it is certainly
possible to model games where player 2 has no idea how many moves player
1 has taken when player 2 is called upon to  move, it is certainly not
typical to focus on the effects of synchrony (and its lack) in games.
On the other hand, in distributed systems, it is typically a major focus.

Byzantine agreement is achievable in certain cases.  Suppose that the
only types of failures are {\em crash failures}---%
a faulty agent behaves according to
the protocol, except that it might crash at some point, after which it
sends no messages.  In the round in which an agent  fails, the agent
sends only a subset of the messages that it is supposed to send
according to its protocol.  Further suppose that the system is
synchronous.  (These two assumptions can be captured by considering the
appropriate context; see \cite[p.~203]{FHMV}.)   In this case, the
following rather simple protocol achieves Byzantine agreement:
\begin{itemize}
\item In the first round, each agent tells every other agent its
initial preference.   
\item For rounds 2 to $t+1$, each agent tells every other agent
everything it has heard in the previous round.  (Thus, for example, in
round 3, agent 1 may tell agent 2 that it heard from agent 3 that
its initial preference was to attack, and that it (agent 3) heard from
agent 2 that its initial preference is to attack, and it heard from
agent 4 that its initial preferences is to retreat, and so on.  This
means that messages get exponentially long, but it is not difficult to
represent this information in a compact way so that the total
communication is polynomial in $n$, the number of agents.)
\item At the end of round $t+1$, if an agent has heard from any other
agent (including itself) that its initial preference was to attack, it
decides to attack; otherwise, it decides to retreat.
\end{itemize}

Why is this correct?  Clearly, if all agents are correct and want to
retreat, then the final decision will be to retreat, since that is the
only preference that other agents hear about (recall that for now we
are considering only crash failures).  Similarly, if all agents
prefer to attack, the final decision will clearly be to attack.  It
remains to show that if some agents prefer to attack and others to
retreat, then all the nonfaulty agents reach the same final decision.
So suppose that $i$ and $j$ are nonfaulty and $i$ decides to attack.
That means that 
$i$ heard that some agent's initial preference was to attack.  If it
heard this first at some round $t' < t+1$, then $i$ will forward this
message to $j$, who will receive it and thus also attack.  On the other
hand, suppose that $i$ heard it first at round $t+1$ in a message from
$i_{t+1}$. Thus, this message must be of the
form ``$i_t$ said at round $t$ that
\ldots that $i_2$ said at round 2 that $i_1$ said at round 1 that its
initial preference was to attack.''  Moreover, the agents $i_1,
\ldots, i_{t+1}$ must all be distinct.  Indeed, it is easy to see that
$i_k$ must crash in round $k$ before sending its message to $i$ (but after
sending its message to $i_{k+1}$), for $k = 1, \ldots, t$, for otherwise
$i$ must have gotten the message from $i_k$, contradicting the
assumption that $i$ first heard at round $t+1$ that some agent's initial
preference was to attack.  Since at
most $t$ agents can crash, it follows that $i_{t+1}$, the agent that
sent the message to $i$, is not faulty, and thus sends the message to
$j$.  Thus, $j$ also decides to attack.  A symmetric argument
shows that if $j$ decides to attack, then so does $i$.

It should be clear that the correctness of this protocol depends on both
the assumptions made: crash failures and synchrony.  Suppose instead that
{\em Byzantine\/} failures are allowed, so that faulty agents can deviate
in arbitrary ways from the protocol; they may ``lie'', send deceiving
messages, and collude to fool the nonfaulty agents in the most malicious
ways.  In this case, the protocol will not work at all.  In fact, it is
known that agreement can be reached in the presence of
Byzantine failures iff $t < n/3$, that is, iff fewer than a third of
the agents can be faulty \cite{PSL}.  The effect of asynchrony is even more
devastating: in an asynchronous system, it is impossible to reach
agreement using a deterministic protocol even if $t=1$ (so that there is at
most one failure) and only crash failures are allowed \cite{FLP}.  The
problem in the asynchronous setting is that if none of the agents have heard
from, say, agent 1, they have no way of knowing whether agent 1 is
faulty or just slow.  Interestingly, there are randomized algorithms
(i.e., behavior strategies) that achieve agreement with arbitrarily high
probability in an asynchronous setting \cite{BenOr,Rab}.  

Finally, note that the protocol above uses $t+1$ rounds.  This bound is
achievable even with
Byzantine failures, provided that that $t < n/3$  \cite{PSL}.  Can we do
better?  In one sense, the answer is no.  Even if only crash failures
are considered, $t+1$ rounds of communication are required in runs where
there are in fact no failures at all \cite{DS}.  
To understand why, 
consider a simple situation where $t=1$, there are only crash failures, 
all agents start with the same initial preference, say to attack, and
there are in fact no failures.  In this case, all the agents can tell each
other in the first round of communication that they want to attack.
Since there are no failures, at the end of the first round, all the
agents will know that all the other agents want to attack.  Thus, they
will know that the ultimate decision  must be to attack, since all the
agents have the same initial preference.  Nevertheless, if they want to
be sure to attack simultaneously, they must wait until the end of the
second round to do so (since $t+1 = 2$ in this case).  

Why is this the case?  Results of Dwork and Moses  \citeyear{DM} give
some insight here.  They  
show that common knowledge among the {\em nonfaulty\/}
agents is necessary and sufficient to attain simultaneous Byzantine
agreement (even though a nonfaulty agent may not know which of the other
agents are faulty).  The nonfaulty agents are what
is called an {\em indexical\/} set in the philosophy literature; a set
whose membership depends on context.  The reason it takes two rounds to
reach agreement even if there are no failures is that, although 
each agent knows that all the other agents had an initial preference
to attack, this fact is not yet common knowledge.  For example, agent 1
might consider it possible that agent 2 was faulty and crashed before
sending a message to agent 3.  In this case, agent 3 would not know that
everyone started with an initial preference to attack.  Moreover, in
this case, agent
3 might consider it possible that agent 2's initial preference was to
retreat, and that agent 2 communicated this preference to agent 1.  
This argument can be extended to show that agent 1 considers it possible
that agent 3 considers it possible that agent 1 considers it possible
\ldots that everyone's initial preference was to retreat.

It might seem that if it takes $t+1$ rounds to reach simultaneous
agreement in the case that there are no failures, then things can only
get worse if there are failures.  However, Dwork and Moses show that this
intuition is misleading.  They use their characterization of agreement
to provide algorithms for  simultaneous Byzantine 
agreement that reach agreement as early as possible, as a function of
the pattern of failures.  Roughly speaking, we can imagine an adversary
with $t$ ``chips'', one for each possible failure.  The adversary plays
a chip by corrupting an agent.  Dwork and Moses' analysis shows that if
the adversary's goal is to make the agreement happen as late as
possible, then the adversary's optimal strategy is, roughly speaking, to
play no more than one chip per round.  If the adversary plays optimally,
agreement cannot be attained before round $t+1$.  Since not corrupting
any agent is an instance of optimal play, it follows that it requires
$t+1$ rounds to reach agreement in runs where there are no failures.
On the other hand, if the adversary plays all $t$ chips in the first
round and none of the faulty agents sends a message, then the correct
agents will know at the end of the first round 
exactly which agents are faulty, and be able to
reach agreement in one more round.  The adversary is best off by keeping the 
agents as uncertain as possible as to which agents are faulty.

Byzantine agreement can be viewed as a game where, at  each step, an
agent can either send a message or decide to attack or retreat.  It is
essentially a game between two teams, the
nonfaulty agents and the faulty agents, whose composition is unknown (at
least by the correct agents).  To model it as a game in the more
traditional sense, we could imagine that the nonfaulty agents are playing
against a new player, the ``adversary''.  One of adversary's moves is 
that of ``corrupting'' an agent: changing its type from ``nonfaulty'' to
``faulty''.   Once an agent is corrupted, what the adversary can do
depends on the failure type
being considered.  In the case of crash failures, the adversary can
decide which of a corrupted agent's  messages will be delivered in the
round in which the agent is corrupted; however, it cannot modify the
messages themselves.  
In the case of Byzantine failures, the adversary essentially gets to
make the moves for agents that have been corrupted; in particular, it can
send arbitrary messages.   

In practice, crash failures occur quite regularly, as a result of
hardware and software failures.  Another failure type considered is
{\em omission failures}.  An agent suffering from an omission failure
behaves according to its protocol, 
except that it may omit to send an arbitrary set of messages
in any given round.  Omission failures are meant to model 
local communications problems (for example, a congested message buffer).
Finally, Byzantine failures represent the worst possible failures,
where we can make no assumption on the behavior of faulty agents.
Byzantine failures are used to capture random behavior on the part
of a system (for example, messages getting garbled in transit),
software errors, and malicious adversaries (for example, hackers).

In the case of crash failures and omission failures (and for Byzantine
failures that are meant to represent random behavior), 
it does not make sense to view the adversary's behavior as strategic,
since in these cases the adversary is not really viewed as having
strategic interests.  However, it would certainly  
make sense, at least in principle, to consider the probability of
failure (i.e., the probability that the adversary corrupts an agent).
But this approach has by and large been
avoided in the literature.  It is very difficult to characterize the
probability distribution of failures over time.  Computer components 
can perhaps be characterized as failing according to an exponential 
distribution (as is done by Babaoglu \citeyear{Bab87}, in one of the few
papers that I am aware of that actually does try to analyze the
situation probabilistically), 
but crash failures can be caused by things other than
component failures (faulty software, for example). Omission failures are
often caused by traffic congestion; as I mentioned before, this is
extremely difficult to characterize probabilistically.  The problems are
even worse when it comes to modeling random Byzantine behavior.

With malicious Byzantine behavior, it may 
well be reasonable to impute strategic behavior to agents (or to an
adversary controlling them). 
However, it is typically very difficult to characterize the payoffs of 
a malicious agent (and, indeed, there is often a great deal of
uncertainty about what a malicious agent's payoffs are).  
The goals of the agents may vary from that of simply trying to delay a
decision to that of causing disagreement.  It is not clear what the
appropriate payoffs should be for attaining these goals.
Thus, the distributed systems literature has chosen to focus instead on
algorithms that are guaranteed to satisfy the specification without
making  assumptions about the adversary's payoffs (or nature's
probabilities, in the case of omission failures and crash failures).  

I believe that some interesting work can be done trying to combine
failures, asynchrony, and strategic incentives.  Some preliminary work
has already been done---for example, Monderer and Tennenholtz 
\citeyear{MT99,MT00} have considered 
timing issues in asynchronous systems, as well as the
structure of the network, and Eliaz \citeyear{Eliaz00} has considered
solution concepts that take failures into account.   However, I believe
that there is much more that can be done.

\section{Specification and Mechanism Design}


Game theory has typically focused on ``small'' games: games that are
easy to describe, such as Prisoner's
Dilemma, Battle of the Sexes, and the Centipede game.  The focus has
been on subtleties regarding  basic issues such as rationality and
coordination.  To the extent that game theory is used to tackle
larger, more practical problems, and especially to the extent that
it is computers, or software agents, playing games, rather than people,
a whole host of new issues arise.  In many cases, the major difficulty may
no longer be conceptual problem of explicating what ought to be
considered ``rational''.  It may be quite obvious what the ``rational'' and
optimal strategy is once we analyze the game.  Rather, the difficulty is
analyzing the game due to its size.  Indeed, part of the difficulty
might even be {\em describing\/} the game.  By way of analogy, $2^{n}
-1$ numbers are needed to describe a probability distribution on a space
characterized by $n$ binary random variables.  For $n = 100$ (not an
unreasonable number in 
practical situations), it is impossible to write down the probability
distribution in the obvious way, let alone do computations with it.
The same issues will surely arise in large games.  
Computer scientists have developed techniques like Bayesian networks for
manipulating probability measures on large spaces \cite{Pearl}; similar
techniques seem applicable to games.  Since these techniques are
discussed in detail by Koller and Milch \citeyear{Koller01} and La Mura
\citeyear{LaMura00}, I do not go into them here.

A related but different problem is involved with dealing with  ``large''
mechanisms.  This, I expect, will be somewhat akin to writing large
programs.  It will be extremely important to specify
carefully exactly what the mechanism must accomplish, and to find
techniques for doing mechanism design in a modular way, so that
mechanisms for solving different problems can be combined in a seamless
way.  

The design and specification of software is well known to be a
critical and often difficult problem in computer science.   The concern
with specification has led to the development of
numerous {\em specification languages}, that is, formal languages 
for expressing carefully the requirements that a protocol must satisfy
(see, for example, \cite{HKT00,Mil,MP1}).%
\footnote{Historically, 
the original work that my colleagues and I did on knowledge and
common knowledge was in part motivated by the desire to find good
tools for designing and specifying distributed protocols.}

It may seem that specification is not so hard.  How hard is it, for
example, to specify a division algorithm?  It gets two inputs $x$ and
$y$ and is supposed to return $x/y$.  Even in this simple
case, there are difficulties.  Are we talking about integer division 
(so that the inputs are integers and the output is an integer, with the
remainder ignored)?  If we are talking about real division, how should the
the answer be represented?  For example, if it is a decimal, what should the
answer be if $x$ and $y$ are 1 and 3, respectively?  How is the infinite
sequence $.333\ldots$ to be represented?  If answers can only be, say,
32 bits long, what happens if $x = 10^{31}$ and $y = 10^{-31}$?  
What should happen if $y=0$?  
And what happens if the inputs are not of the right type (i.e., they are
letters instead of numbers)?  The key point is that a good specification
will need to describe what should happen in all the ``unexpected'' cases. 

This can get particularly difficult once we try to take into account
failures and asynchrony.  Imagine trying to specify 
a good mechanism for a distributed auction.  The specification will need
to take into account the standard distributed concerns of 
asynchronous communication, failures, garbled communication, economic
issues like failure to pay, and strategic issues (including 
strategic uses of computing difficulties, such as  pretending not to
have received messages or to have received them late).  Thus, its
specification must address what should happen if a process fails in the
middle of transmitting a bid, how to deal with agents bidding on slow
lines, and so on.

\commentout{
Specifying desired behavior in the presence of failures is 
even trickier.  Consider the informal specification that I gave of Byzantine
agreement in the preceding section.  One part of the specification says 
``all nonfaulty soldiers reach the same decision''.  This could mean
\begin{itemize}
\item[(a)] If $i$ and $j$ are nonfaulty throughout run $r$, then if $i$
decides on $y$, then so does $j$.
\item[(b)] If $i$ is nonfaulty at the point $(r,m)$ and has decided $y$
by $(r,m)$ (that is, if $i$ decided on $y$ in run $r$ at some round $m'
\le m$) and $j$ is nonfaulty at $(r,m')$ and has decided $y'$ by
$(r,m')$, then  
$y=y'$. 
\item[(c)] If $i$ is nonfaulty at $(r,m)$, and $i$ is about
decide $y$ at $(r,m)$ (that is, $i$ has not decided before $(r,m)$, and
$i$'s protocol dictates that $i$ should decide $y$ at the point
$(r,m)$), and $j$ is nonfaulty at $(r,m')$ and is about to 
decide $y'$, then $y = y'$.
\end{itemize}
It is not hard to show that (c) implies (b) and that (b) implies (a),
but, in general, (a), (b), and (c) are not equivalent.
These subtle distinctions can make a big difference in what counts as an
acceptable Byzantine agreement algorithm.
}

Things get significantly more complicated if we try to specify notions
like security.  What exactly does it mean that a mechanism is secure?  
What types of attacks can be tolerated?  For example, how should the
mechanism behave if there is a denial-of-service attack?  
I suspect that questions regarding security and fault-tolerance will
turn out to be closely intertwined with strategic issues.  Thus,
finding appropriate techniques for specifying mechanisms will not simply
be a matter of lifting standard techniques from software specification.

\section{Conclusions}
I have focused on one set of
issues at the interface of computer science and game theory here, which
arise from work in distributed computing.   
As I hope this discussion  has made clear, I think that game theorists
need to take more seriously issues like fault tolerance, asynchrony, the
representation of knowledge and uncertainty, the difficult in the design
and analysis of large mechanisms and games, and problems of
specification.  On the other hand, 
I think computer scientists need to take strategic concerns more
seriously in the design and analysis of distributed protocols.
These issues are not just of theoretical interest.
They arise, for example, when we consider the 
design of Internet agents.  We will certainly need to take into account
failures, and no company would want to claim to support software for
agents that bid in auctions  that has not been carefully specified.%
\footnote{The correctness of the agents will also have to be {\em verified\/}
somehow.  Verification is yet another issue of great concern in computer
science that may prove relevant to game theory.  Much work has gone into
finding (preferably automatic or semi-automatic) techniques to check that 
a protocol satisfies a specification (see, for example,
\cite{AptOlderog91,CGP99}).}
The specification of the agents will, in turn, depend on a careful
specification of the mechanism in which they will be participating.

These issues represent only part of the commonality in interests that I
see between computer science and game theory.  I have already hinted at
another important area of commonality: that of finding compact
representations of games. Other issues of common interest include
learning, mental-level modeling of beliefs \cite{BraTenjour}, qualitative
decision theory (see the bibliography of over 290 papers at {\tt
http:/$\!$/www.medg.lcs.mit.edu/qdt/bib/unsorted.bib}).
With the growing awareness of the commonality between computer science
and game theory, I look forward to a great deal of fruitful interaction
between the fields in the coming years.

\subsection*{Acknowledgments}  I would like to thank Ron Fagin, Yoram
Moses, Ariel Rubinstein, Moshe Tennenholtz, and Moshe Vardi for useful
comments on an earlier draft of this paper.

\bibliographystyle{chicago}

\begin{thebibliography}{}

\bibitem[\protect\citeauthoryear{Apt and Olderog}{Apt and
  Olderog}{1991}]{AptOlderog91}
Apt, K.~R. and E.-R. Olderog (1991).
\newblock {\em Verification of sequential and concurrent programs}.
\newblock New York: Springer-Verlag.

\bibitem[\protect\citeauthoryear{Aumann}{Aumann}{1976}]{Au}
Aumann, R.~J. (1976).
\newblock Agreeing to disagree.
\newblock {\em Annals of Statistics\/}~{\em 4\/}(6), 1236--1239.

\bibitem[\protect\citeauthoryear{Babaoglu}{Babaoglu}{1987}]{Bab87}
Babaoglu, O. (1987).
\newblock On the reliability of consensus-based fault-tolerant distributed
  computing systems.
\newblock {\em ACM Transactions on Computer Systems\/}~{\em 5}, 394--416.

\bibitem[\protect\citeauthoryear{{Ben-Or}}{{Ben-Or}}{1983}]{BenOr}
{Ben-Or}, M. (1983).
\newblock Another advantage of free choice: completely asynchronous agreement
  protocols.
\newblock In {\em Proc.~2nd ACM Symp.~on Principles of Distributed Computing},
  pp.\  27--30.

\bibitem[\protect\citeauthoryear{Brafman and Tennenholtz}{Brafman and
  Tennenholtz}{1997}]{BraTenjour}
Brafman, R.~I. and M.~Tennenholtz (1997).
\newblock Modeling agents as qualitative decision-makers.
\newblock {\em Artificial Intelligence\/}~{\em 94}, 217--268.

\bibitem[\protect\citeauthoryear{Chor and Dwork}{Chor and Dwork}{1989}]{CD89}
Chor, B. and C.~Dwork (1989).
\newblock Randomization in {Byzantine} agreement.
\newblock In {\em Advances in Computing Research 5: Randomness and
  Computation}, pp.\  443--497. JAI Press.

\bibitem[\protect\citeauthoryear{Clarke, Grumberg, and Peled}{Clarke
  et~al.}{1999}]{CGP99}
Clarke, E.~M., O.~Grumberg, and D.~A. Peled (1999).
\newblock {\em Model Checking}.
\newblock Cambridge, Mass.: MIT Press.

\bibitem[\protect\citeauthoryear{Dolev and Strong}{Dolev and Strong}{1982}]{DS}
Dolev, D. and H.~R. Strong (1982).
\newblock Requirements for agreement in a distributed system.
\newblock In H.~J. Schneider (Ed.), {\em Distributed Data Bases}, pp.\
  115--129. Amsterdam: North-Holland.

\bibitem[\protect\citeauthoryear{Dwork and Moses}{Dwork and Moses}{1990}]{DM}
Dwork, C. and Y.~Moses (1990).
\newblock Knowledge and common knowledge in a {B}yzantine environment: crash
  failures.
\newblock {\em Information and Computation\/}~{\em 88\/}(2), 156--186.

\bibitem[\protect\citeauthoryear{Eliaz}{Eliaz}{2000}]{Eliaz00}
Eliaz, K. (2000).
\newblock Fault-tolerant implementation.
\newblock Unpublished manuscript.

\bibitem[\protect\citeauthoryear{Fagin, Halpern, Moses, and Vardi}{Fagin
  et~al.}{1995}]{FHMV}
Fagin, R., J.~Y. Halpern, Y.~Moses, and M.~Y. Vardi (1995).
\newblock {\em Reasoning about Knowledge}.
\newblock Cambridge, Mass.: MIT Press.

\bibitem[\protect\citeauthoryear{Fischer}{Fischer}{1983}]{Fisbyz}
Fischer, M.~J. (1983).
\newblock The consensus problem in unreliable distributed systems.
\newblock Technical Report RR-273, Yale University.
\newblock Also appears in {\em Foundations of Computation Theory}, ed. M.
  Karpinski, Lecture Notes in Computer Science, Vol. 185, Springer Verlag,
  1983, pp. 127--140.

\bibitem[\protect\citeauthoryear{Fischer, Lynch, and Paterson}{Fischer
  et~al.}{1985}]{FLP}
Fischer, M.~J., N.~A. Lynch, and M.~S. Paterson (1985).
\newblock Impossibility of distributed consensus with one faulty processor.
\newblock {\em Journal of the ACM\/}~{\em 32\/}(2), 374--382.

\bibitem[\protect\citeauthoryear{Gray}{Gray}{1978}]{Gray}
Gray, J. (1978).
\newblock Notes on database operating systems.
\newblock In R.~Bayer, R.~M. Graham, and G.~Seegmuller (Eds.), {\em Operating
  Systems: An Advanced Course}, Lecture Notes in Computer Science, Vol. 66.
  Berlin/New York: Springer-Verlag.
\newblock Also appears as IBM Research Report RJ 2188, 1978.

\bibitem[\protect\citeauthoryear{Halpern}{Halpern}{1995}]{Hal32}
Halpern, J.~Y. (1995).
\newblock Reasoning about knowledge: a survey.
\newblock In D.~M. Gabbay, C.~J. Hogger, and J.~A. Robinson (Eds.), {\em
  Temporal and Epistemic Reasoning}, Volume~4 of {\em Handbook of of Logic in
  Artificial Intelligence and Logic Programming}, pp.\  1--34. Oxford, U.K.:
  Oxford University Press.

\bibitem[\protect\citeauthoryear{Halpern}{Halpern}{1997}]{Hal15}
Halpern, J.~Y. (1997).
\newblock On ambiguities in the interpretation of game trees.
\newblock {\em Games and Economic Behavior\/}~{\em 20}, 66--96.

\bibitem[\protect\citeauthoryear{Halpern and Fagin}{Halpern and
  Fagin}{1989}]{HF87}
Halpern, J.~Y. and R.~Fagin (1989).
\newblock Modelling knowledge and action in distributed systems.
\newblock {\em Distributed Computing\/}~{\em 3\/}(4), 159--179.
\newblock A preliminary version appeared in {\em Proc.~4th ACM Symposium on
  Principles of Distributed Computing}, 1985, with the title ``A formal model
  of knowledge, action, and communication in distributed systems: preliminary
  report''.

\bibitem[\protect\citeauthoryear{Halpern and Moses}{Halpern and
  Moses}{1990}]{HM1}
Halpern, J.~Y. and Y.~Moses (1990).
\newblock Knowledge and common knowledge in a distributed environment.
\newblock {\em Journal of the ACM\/}~{\em 37\/}(3), 549--587.
\newblock A preliminary version appeared in {\em Proc.~3rd ACM Symposium on
  Principles of Distributed Computing}, 1984.

\bibitem[\protect\citeauthoryear{Harel, Kozen, and Tiuryn}{Harel
  et~al.}{2000}]{HKT00}
Harel, D., D.~C. Kozen, and J.~Tiuryn (2000).
\newblock {\em Dynamic Logic (Foundations of Computing)}.
\newblock Cambridge, Mass.: MIT Press.

\bibitem[\protect\citeauthoryear{Hintikka}{Hintikka}{1962}]{Hi1}
Hintikka, J. (1962).
\newblock {\em Knowledge and Belief}.
\newblock Ithaca, N.Y.: Cornell University Press.

\bibitem[\protect\citeauthoryear{Kfir-Dahav, Monderer, and
  Tennenholtz}{Kfir-Dahav et~al.}{2000}]{KMT00}
Kfir-Dahav, N.~E., D.~Monderer, and M.~Tennenholtz (2000).
\newblock Mechanism design for resource bounded agents.
\newblock Unpublished manuscript.

\bibitem[\protect\citeauthoryear{Koller and Milch}{Koller and
  Milch}{2001}]{Koller01}
Koller, D. and B.~Milch (2001).
\newblock Structured models for multiagent interactions.
\newblock In {\em Theoretical Aspects of Rationality and Knowledge:
  Proc.~Eighth Conference (TARK2001)}, pp.\  233--248. San Francisco, Calif.:
  Morgan Kaufmann.

\bibitem[\protect\citeauthoryear{{La Mura}}{{La Mura}}{2000}]{LaMura00}
{La Mura}, P. (2000).
\newblock Game networks.
\newblock In {\em Proc.~Sixteenth Conference on Uncertainty in Artificial
  Intelligence (UAI 2000)}.

\bibitem[\protect\citeauthoryear{Linial}{Linial}{1994}]{Linial94}
Linial, N. (1994).
\newblock Games computers play: game-theoretic aspects of computing.
\newblock In R.~J. Aumann and S.~Hart (Eds.), {\em Handbook of Game Theory with
  Economic Applications}, Volume~II, pp.\  1340--1395. Amsterdam:
  North-Holland.

\bibitem[\protect\citeauthoryear{Manna and Pnueli}{Manna and
  Pnueli}{1992}]{MP1}
Manna, Z. and A.~Pnueli (1992).
\newblock {\em The Temporal Logic of Reactive and Concurrent Systems},
  Volume~1.
\newblock Berlin/New York: Springer-Verlag.

\bibitem[\protect\citeauthoryear{Milner}{Milner}{1980}]{Mil}
Milner, R. (1980).
\newblock {\em A Calculus of Communicating Systems}.
\newblock Lecture Notes in Computer Science, Vol. 92. Berlin/New York:
  Springer-Verlag.

\bibitem[\protect\citeauthoryear{Monderer and Tennenholtz}{Monderer and
  Tennenholtz}{1999a}]{MT99}
Monderer, D. and M.~Tennenholtz (1999a).
\newblock Distributed games.
\newblock {\em Games and Economic Behavior\/}~{\em 28}, 55--72.

\bibitem[\protect\citeauthoryear{Monderer and Tennenholtz}{Monderer and
  Tennenholtz}{1999b}]{MT00}
Monderer, D. and M.~Tennenholtz (1999b).
\newblock {Distributed Games: From Mechanisms to Protocols}.
\newblock In {\em AAAI-99}, pp.\  32--37.

\bibitem[\protect\citeauthoryear{Neyman}{Neyman}{1985}]{Ney85}
Neyman, A. (1985).
\newblock Bounded complexity justifies cooperation in finitely repated
  prisoner's dilemma.
\newblock {\em Economic Letters\/}~{\em 19}, 227--229.

\bibitem[\protect\citeauthoryear{Pearl}{Pearl}{1988}]{Pearl}
Pearl, J. (1988).
\newblock {\em Probabilistic Reasoning in Intelligent Systems}.
\newblock San Francisco, Calif.: Morgan Kaufmann.

\bibitem[\protect\citeauthoryear{Pease, Shostak, and Lamport}{Pease
  et~al.}{1980}]{PSL}
Pease, M., R.~Shostak, and L.~Lamport (1980).
\newblock Reaching agreement in the presence of faults.
\newblock {\em Journal of the ACM\/}~{\em 27\/}(2), 228--234.

\bibitem[\protect\citeauthoryear{Piccione and Rubinstein}{Piccione and
  Rubinstein}{1997}]{PR97}
Piccione, M. and A.~Rubinstein (1997).
\newblock On the interpretation of decision problems with imperfect recall.
\newblock {\em Games and Economic Behavior\/}~{\em 20\/}(1), 3--24.

\bibitem[\protect\citeauthoryear{Rabin}{Rabin}{1983}]{Rab}
Rabin, M.~O. (1983).
\newblock Randomized {B}yzantine generals.
\newblock In {\em Proc.~24th IEEE Symp.~on Foundations of Computer Science},
  pp.\  403--409.

\bibitem[\protect\citeauthoryear{Rubinstein}{Rubinstein}{1986}]{Rub85}
Rubinstein, A. (1986).
\newblock Finite automata play the repeated prisoner's dilemma.
\newblock {\em Journal of Economic Theory\/}~{\em 39}, 83--96.

\bibitem[\protect\citeauthoryear{Rubinstein}{Rubinstein}{1989}]{Rub89}
Rubinstein, A. (1989).
\newblock The electronic mail game: strategic behavior under ``almost common
  knowledge''.
\newblock {\em American Economic Review\/}~{\em 79}, 385--391.

\bibitem[\protect\citeauthoryear{Yemini and Cohen}{Yemini and Cohen}{1979}]{YC}
Yemini, Y. and D.~Cohen (1979).
\newblock Some issues in distributed processes communication.
\newblock In {\em Proc.~of the 1st International Conf. on Distributed Computing
  Systems}, pp.\  199--203.

\end{thebibliography}

\end{document}